\documentclass[twocolumn]{aastex631}
\usepackage[utf8]{inputenc}
\usepackage{mathtools}
\usepackage{nameref}
\usepackage{wasysym}
\usepackage{makecell}
\usepackage[caption=false]{subfig}
\usepackage{hyperref}
\usepackage[capitalise]{cleveref}
\usepackage{booktabs}
\usepackage{float}
\usepackage{blindtext}
\usepackage{enumitem}
\usepackage{xcolor}

\graphicspath{{images/}}

\received{November 8, 2023}
\revised{March 15, 2024}
\accepted{March 21, 2024}
\submitjournal{ApJ}

\shorttitle{Can Jupiter's atmospheric metallicity be different from the deep interior?}
\shortauthors{Simon Müller and Ravit Helled}

\NewPageAfterKeywords{}
\turnoffedit

\newcommand{\addone}[1]{#1}
\newcommand{\addtwo}[1]{#1}

\begin{document}

\title{Can Jupiter's atmospheric metallicity be different from the deep interior?} 

\author[0000-0002-8278-8377]{Simon Müller}
\affiliation{Department of Astrophysics, University of Zürich, \\
             Winterthurerstrasse 190, 8057 Zürich, Switzerland}
\email{simonandres.mueller@uzh.ch}

\author[0000-0001-5555-2652]{Ravit Helled}
\affiliation{Department of Astrophysics, University of Zürich, \\
             Winterthurerstrasse 190, 8057 Zürich, Switzerland}
             
\correspondingauthor{Simon Müller}

\begin{abstract}
\addone{Updated formation and structure models of Jupiter predict a metal-poor envelope. This is at odds with the two to three times solar metallicity measured by the Galileo probe. Additionally, Juno data imply that water and ammonia are enriched. Here we explore whether Jupiter can have a deep radiative layer separating the atmosphere from the deeper interior. The radiative layer could be caused by a hydrogen-transparency window or depletion of alkali metals.
We show that heavy-element accretion during Jupiter's evolution can lead to the desired atmospheric enrichment and that this configuration is stable over billions of years. The origin of the heavy elements could be cumulative small impacts or one large impact. The preferred scenario requires a deep radiative zone due to a local reduction of the opacity \addtwo{at $\sim$ 2000 K by $\sim$ 90\%}, which is supported by Juno data, and vertical mixing through the boundary with a similar efficiency to molecular diffusion \addtwo{($D \lesssim 10^{-2}$ cm$^2$/s)}. Therefore, most of Jupiter's molecular envelope could have solar composition while its uppermost atmosphere is enriched with heavier elements. The enrichment likely originates from the accretion of solid objects. This possibility resolves the long-standing mismatch between Jupiter's interior models and atmospheric composition measurements. Furthermore, our results imply that the measured atmospheric composition of exoplanets does not necessarily reflect their bulk compositions. 
We also investigate whether the enrichment could be due to the erosion of a dilute core and show that this is highly unlikely. The core-erosion scenario is inconsistent with evolution calculations, the deep radiative layer, and published interior models.}
\end{abstract}

\keywords{planets and satellites: evolution, gaseous planets, atmospheres, interiors, composition --- methods: numerical}

\section{Introduction}\label{sec:introduction}

Constraining Jupiter's bulk composition and internal structure is crucial for understanding its formation and evolution history. Jupiter's interior is typically constrained using gravity data along with other measurements of its basic properties \cite[e.g.,][and references therein]{2022arXiv220210046H}. Our understanding of Jupiter's interior has significantly improved thanks to the Juno mission \citep[e.g.,][]{2017Sci...356..821B,2020GeoRL..4786572D}. We now know that Jupiter is likely to have a complex interior that is inhomogeneous with, possibly, a fuzzy core \citep[e.g.,][]{Wahl2017,2021PSJ.....2..241N,2022PSJ.....3...89I,2022A&A...662A..18M,2022PSJ.....3..185M,2023arXiv230209082H}. These recent structure models typically require a low-density atmosphere to match the gravitational moments and often favor a sub-solar metallicity in the envelope.  

Another important property that can constrain Jupiter's interior is its atmospheric composition. Measurements from the Galileo probe and the Juno mission show that Jupiter's atmosphere has a metallicity of about two to three times solar \citep[e.g.,][]{2000JGR...10515061M,2020NatAs...4..609L}. This is at odds with recent structure models of Jupiter, \addone{which creates a tension between interior models and observations.} It should be kept in mind that these measurements probed a shallow region of Jupiter's atmosphere of about one to thirty bar. If most of Jupiter's envelope is convective as is commonly assumed, enriching the envelope would require a significant amount of heavy elements on the order of a few Earth masses. \addone{While there are formation pathways to enrich the envelope via the accretion of super-solar gas \citep{2022PSJ.....3..141A} or solids during or after the runaway gas accretion stage \citep{2022ApJ...926L..37S,2023MNRAS.519.1713S}, at the moment no formation model can lead to an overall enrichment of Jupiter that is about three times solar. As a result, from a formation perspective, Jupiter's envelope is expected to have solar composition \citep[e.g.,][and references therein]{2021exbi.book...12H,2022Icar..37814937H}.}

\addone{The tension between interior models and atmospheric measurements} raises the possibility that the metallicity in the upper atmosphere of Jupiter is higher than the metallicity of the envelope. It is therefore currently unclear whether the measured atmospheric composition is a good representation of the bulk of the envelope. \addone{Jupiter may harbor} a deep radiative zone that disconnects the outer envelope from the deeper envelope. 

\addone{Such a radiative zone in Jupiter was first suggested by \citet{1994Icar..112..354G,1994Icar..112..337G}. Using their at the time state-of-the-art opacity calculations, \citet{1994Icar..112..337G} found that there is a dip in the hydrogen opacity around 1000 to 2000 K, resulting in a sub-adiabatic radiative temperature gradient. The contribution of metals (Na, Mg, and others) and H$_2$O significantly increased the opacity, but still allowed for a radiative window. Using these opacities, \citet{1994Icar..112..354G} constructed non-adiabatic interior models of Jupiter, which included a deep radiative zone surrounded by two convective zones. More recent opacity calculations \citep{Freedman2008}, however, found that the opacity in the local minimum range is much larger than considered by \citet{1994Icar..112..354G,1994Icar..112..337G}. This is because there is a significant contribution of the pressure-broadened alkali metal opacity in this region, resulting in a sufficiently high enough absorption to prevent substantial transport of energy by radiation. This was previously also suggested by \citet{Guillot2004} in a re-evaluation of their earlier work.}

\addone{Such high opacities, may not apply to Jupiter and in fact, there are new indications that Jupiter could have a deep radiative layer.  Recently, \citet{2023NatAs...7..678C} found evidence of a deep radiative zone by comparing their models to Juno microwave radiometer data. In another study,  \citet{2023ApJ...952L..27B} argued that alkali metals could be significantly depleted in Jupiter,  leading to a significant opacity reduction and the existence of a radiative layer. Indeed, it was shown in \citet{Freedman2008} that without the presence of alkali metals, the radiative opacity around 2000 K at a density of $10^{-2}$ g/cm$^3$ is reduced by about a factor of twenty, which corresponds to a reduction of about 95\% in the opacity. The inferred depletion of alkali metals could therefore restore the original idea of \citet{1994Icar..112..354G,1994Icar..112..337G} of a deep radiative zone.}

\begin{figure}[ht]
    \centering
    \includegraphics[width=\columnwidth]{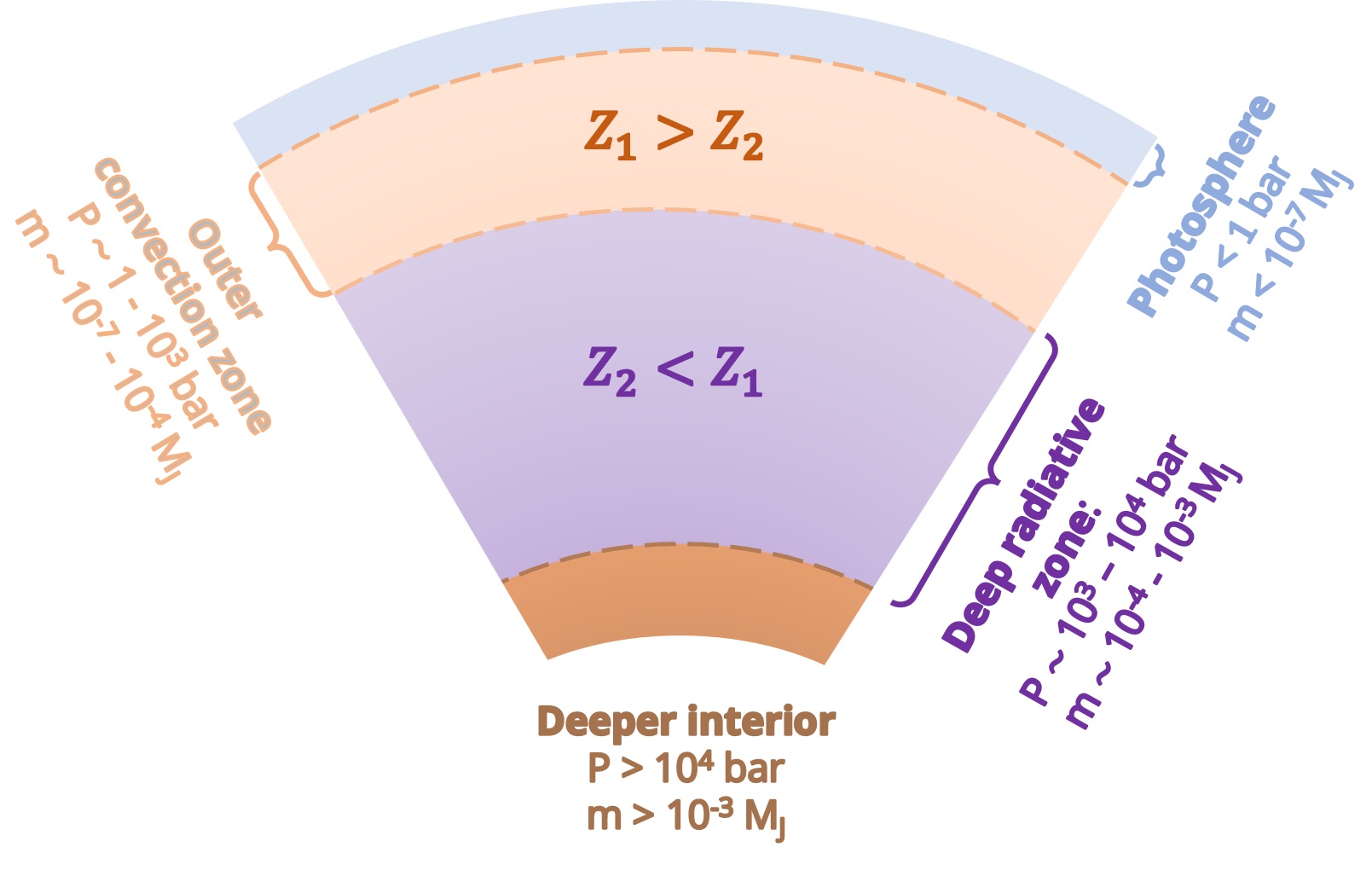}
    \caption{A sketch of Jupiter's structure showing the radiative-convective-radiative layering presented in this work. The outermost layer down to $\sim$ 1 bar is the radiative photosphere and sits on top of an outer convection zone ($\sim 1 - 10^3$ bar) with heavy-element mass fraction $Z_1$. Below that is a deep radiative zone, between $\sim 10^3 - 10^4$ bar, with a heavy element mass fraction $Z_2 < Z_1$. The deeper interior harbors a convection zone, which is needed to generate the magnetic field, and possibly a dilute core that may be stably stratified. The mass coordinate is defined from the outside-in, with $m = 0$ at the surface and 1 M$_J$ at the center.}
    \label{fig:introduction_pie}
\end{figure}

Consequently, there may yet be radiative-convective-radiative layering in the upper envelope of Jupiter. This would allow for a decrease of heavy elements with depth, as discussed above. The resulting structure is sketched in \cref{fig:introduction_pie}. At pressures below about one bar, there is a radiative \textit{photosphere}. Farther down, there is an \textit{outer convection zone}, extending between about 1-1,000 bar, with metallicity $Z_1$. Next is the suggested \textit{deep radiative zone} with a lower metallicity $Z_2 < Z_1$, between about 1,000-10,000 bar. Deeper still is the rest of the envelope, consisting mostly of molecular and metallic hydrogen, and a dilute/compact core. The viability of this model was recently investigated by \citet{2023A&A...680L...2H}. While they find that such a structure could be stable, they suggest it is somewhat unlikely in their specific setup of Jupiter's evolution. However, the evolution models used in \citet{2023A&A...680L...2H} considered a narrow range of parameters and did not fully explore the potential pathways that would lead to an enriched atmosphere.

In the above picture, to explain the super-solar atmospheric metallicity measurements, only the outer convection zone needs to be enriched. The advantage of this scenario is that very little mass is required to explain the observed metallicity: The entire outer convection zone contains $\sim10^{-4}$ of Jupiter's mass ($\sim 10^{-2} M_\oplus)$. Raising its metallicity from one to three times solar would require only about $\sim 10^{-4} M_\oplus$. The impact of the comet Shoemaker-Levy 9 has shown that Jupiter is still accreting large solid objects even today. \addone{While some estimates show that the impact rate of similar objects on Jupiter today could be one every few hundred years \citep{1997Icar..126..138R}, the rate of impacts is very uncertain  \citep[e.g.,][]{2018A&A...617A..68H}. Shoemaker-Levy 9 had a mass of $\sim 10^{-12} M_\oplus$ \citep{2004jpsm.book..159H}, the accretion of similar objects would have had to occur at a rate of one every ten years over the lifetime of Jupiter. The impact flux was much higher in the past, and the accreted objects were likely more massive \citep[e.g.,][]{Liu2019,2023PSJ.....4..168B}. It is therefore plausible that similar objects could provide the needed atmospheric enrichment.}

In this study, we investigate the potential pathways to enrich Jupiter's atmosphere. We simulated  Jupiter's evolution using a one-dimensional thermal evolution model as described in \S \ref{sec:methods}, with a focus on the opacities and the mixing of chemical elements. We considered several scenarios: In \S \ref{sec:radiative_zone} we investigate the appearance and evolution of a deep radiative zone in Jupiter. Then, in \S \ref{sec:enrichment_from_above}, we present models where the outer convection zone is enriched from above by the accretion of heavy-element-rich objects. In \S \ref{sec:enrichment_from_below} we examine whether Jupiter's envelope could be enriched from below by eroding a primordial dilute core. We discuss our results in \S \ref{sec:discussion}, and present our conclusions in \S \ref{sec:conclusions}.

\section{Methods}\label{sec:methods}

To model Jupiter's evolution, we used a modified version of the Modules for Experiments in Stellar Astrophysics code (MESA; \citet{Paxton2011,Paxton2013,Paxton2015,Paxton2018,Paxton2019,2022arXiv220803651J}). MESA uses the Henyey method \citep{Henyey1965} to solve the equations of stellar and planetary structure \citep[e.g.,][]{Kippenhahn2012}. We update the models from \citet{2020ApJ...903..147M,Mueller2020} with a new hydrogen-helium equation of state which includes non-ideal interactions \citep{2021ApJ...917....4C}, since they are important for interior and evolution models of giant planets \citep{2023arXiv230207902H}. \addone{The starting point of our models is after Jupiter has formed and the solar nebula has disappeared.} In investigating the enrichment of the atmosphere, both the opacity and the mixing of chemical elements are crucial ingredients as discussed below.

\subsection{Conductive and radiative opacities}\label{sec:methods_opacities}

In the deeper interior at high temperatures and pressures, the conductive opacity is dominant. We use the default conductive opacities in MESA, which are an extended version of the results from \citet{Cassisi2007} privately communicated by A.Y. Potekhin. In the outer envelope where temperatures are low (below a few thousand K), we use the radiative opacities from \citet{Freedman2014} (hereafter Freedman opacity) as a baseline. \citet{1994Icar..112..354G,1994Icar..112..337G} suggested that Jupiter could have a radiative zone in its outer envelope due to a transparency window of hydrogen. While the opacities from \citet{Freedman2014} show a similar transparency window, it is not transparent enough to create a radiative zone. However, recent thermo-chemical and diffusion models lend support to the idea of a radiative zone in the outer envelope (\citet{2023NatAs...7..678C}; see also \S \ref{sec:discussion} for a discussion). Therefore, here we modify the Freedman opacity such that the hydrogen-transparency window is deeper and could lead to a radiative zone.

\begin{figure}[ht]
    \centering
    \includegraphics[width=\columnwidth]{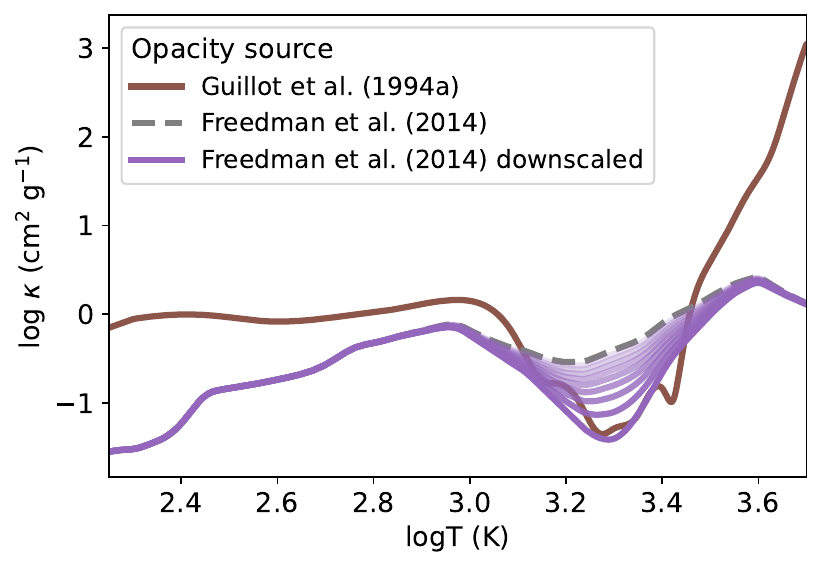}
    \caption{Opacity as a function of temperature for a present-day Jupiter model. The solid brown line shows the opacity calculated with the table from \citet{1994Icar..112..354G}. The dashed gray line shows the \citet{Freedman2014} opacity, while the purple lines show the downscaled versions for different opacity scaling factors \addone{$\alpha = 0.1$ to $0.9$ in steps of $0.1$ from as described by Eq. \ref{eq:opacity_scaling}}}.
    \label{fig:methods_opacity}
\end{figure}

To mimic the dip in the opacity at $\sim$ 2,000 K from \citet{1994Icar..112..354G}, we reduce the Freedman radiative opacity $\kappa_f$ as follows:

\begin{equation}
    \kappa_r = \kappa_f \left(1 - \alpha \, \textrm{e}{^{-0.5 (\log T - \mu) / \sigma}}\right) \, ,
    \label{eq:opacity_scaling}
\end{equation}

where $\kappa_r$ is the radiative opacity that is used for the evolution calculation, and $\log T$ is the logarithm of the local temperature in K.  We set $\mu = 3.3$ and $\sigma = 0.15$ to qualitatively match the location and width of the opacity dip. Additionally, the opacity scaling factor $\alpha$ parameterizes the depth of the dip. The resulting opacity for a present-day interior model of Jupiter is shown in \cref{fig:methods_opacity}. Compared to \addone{the opacities from \citet{1994Icar..112..337G}}, the Freedman opacity is smaller at high and lower temperatures and only has a small dip around 2,000 K. Using $\alpha = 0.9$, we found a good qualitative agreement in the depth of the dip and therefore we used this value in our nominal models. \addone{We note that such a parametric modification to the opacity is a simplification, and indeed \citet{2023A&A...680L...2H} suggest that the opacity from \citet{1994Icar..112..337G} may need to be increased by a factor of a few for interior models with a deep radiative zone. However, if alkali metals are indeed depleted in Jupiter's atmosphere, the opacity could be significantly reduced (by a factor of twenty). Therefore, a reduction of the Freedman opacity by 90\% is reasonable and is taken to be the nominal case in this study.}

\addone{In \S \ref{sec:radiative_zone} we present evolution models using this value of $\alpha$ and investigate the appearance of the radiative zone. Additional models that investigate the dependency on $\alpha$ are presented in Appendix \ref{sec:appendix_kippenhahn_diagrams}.}

\subsection{Mixing of chemical elements}\label{sec:methods_mixing}

In evolution models, the mixing of chemical elements by large-scale convection is modeled as diffusive processes in the mixing-length theory framework \citep[e.g.,][]{Kippenhahn2012}. We use the Ledoux criterion \citep{1947ApJ...105..305L} to determine whether a region is convective or radiative. There is also the possibility of double-diffusive convection in regions with a non-zero mean-molecular weight gradient. This can lead to semi- or thermohaline convection \citep[e.g.,][]{Wood2013,Radko2014,2018AnRFM..50..275G}, depending on whether the mean-molecular weight increases or decreases radially towards the center of the planet. In our models, semi- and thermohaline convection are implemented as described in \citet{Langer1983,Langer1985} and \citet{2013ApJ...768...34B}. \addone{In particular, the condition for linear stability against thermohaline instabilities is $R_0 \geq \tau^{-1}$, where $R_0 \equiv (\nabla_r - \nabla_{ad}) / B$ is the density parameter and $\tau \equiv K_\mu / K_T$ is the ratio of composition and thermal diffusivity. Here, $\nabla_r$ and $\nabla_{ad}$ are the radiative and adiabatic temperature gradients, $B$ is the composition term from the Ledoux criterion, and $K_\mu$ and $K_T$ are the composition and thermal diffusivity, respectively. If $R_0 < \tau^{-1}$, then the temperature gradient and the diffusion coefficient are adjusted to allow for thermohaline mixing \cite[see][for details]{2013ApJ...768...34B}.}

For the models where the atmosphere is enriched from above, it is also possible that non-convective diffusion plays an important role, in particular, because the masses that are involved are small. To account for the vertical transport of chemical species in the absence of other instabilities (large-scale, semi-, and thermohaline convection), we set the diffusion coefficient to some minimal value $D_{min}$ that is non-zero even in the stable, purely radiative regions. 

The diffusion coefficient deeper in the atmosphere at several thousand K is poorly constrained. In recent work where a deep radiative zone in Jupiter was suggested, the diffusion coefficient was set to $1$ cm$^2$ s$^{-1}$ somewhat ad-hoc \citep{2023NatAs...7..678C}. An alternative estimate can be derived from the molecular diffusion of two species for an ideal gas. Assuming that the two species are hydrogen and water, the molecular diffusion coefficient is given by:

\begin{equation}
    D \simeq 0.02 \, \textrm{cm}^{2} \, \textrm{s}^{-1} \left(\frac{T}{500 \, \textrm{K}}\right)^{3/2} \left(\frac{p}{100 \,  \textrm{bar}}\right)^{-1} \, ,
    \label{eq:molecular_diffusion_coefficient}
\end{equation}

where $T$ and $p$ are the temperature in K and pressure in bar. At the proposed location of the deep radiative zone, this would yield a diffusion coefficient of $D \simeq 10^{-3}$ cm$^2$ s$^{-1}$ \addone{which would lead to vertical transport of chemical species even if the region is fully stable against convection, semi- or thermohaline convection}. Given the uncertain diffusion coefficient, we investigate for which values of $D_{min}$ the vertical transport across a stable region would be sufficient to smooth out any composition gradients over relevant timescales completely. However, our nominal models use the estimate for molecular diffusion. We further note chemical elements likely diffuse at different rates. For simplification, the heavy elements are represented by a single chemical species.

\section{A deep radiative zone in Jupiter}\label{sec:radiative_zone}

In this section, we investigate the appearance of the radiative zone as Jupiter cools due to the modified radiative opacities (see \S \ref{sec:methods_opacities}). Our initial post-formation models assume a hot-start for Jupiter, which is supported by formation models as well as direct-imaging observations \citep{2013A&A...558A.113M,Berardo2017a,Berardo2017b, Cumming2018,Marleau2017,2019ApJ...881..144M,2019ApJ...878L..37F}. To simplify our initial model, we also assume a core-envelope structure. Indeed, if a composition gradient were present, the cooling would be affected. However, while it would change the exact timing of when the radiative zone appears or disappears, it would not change the general result. We then let the planet cool down and tracked how the radiative zone(s) evolved.

\begin{figure}[ht]
    \centering
    \includegraphics[width=\columnwidth]{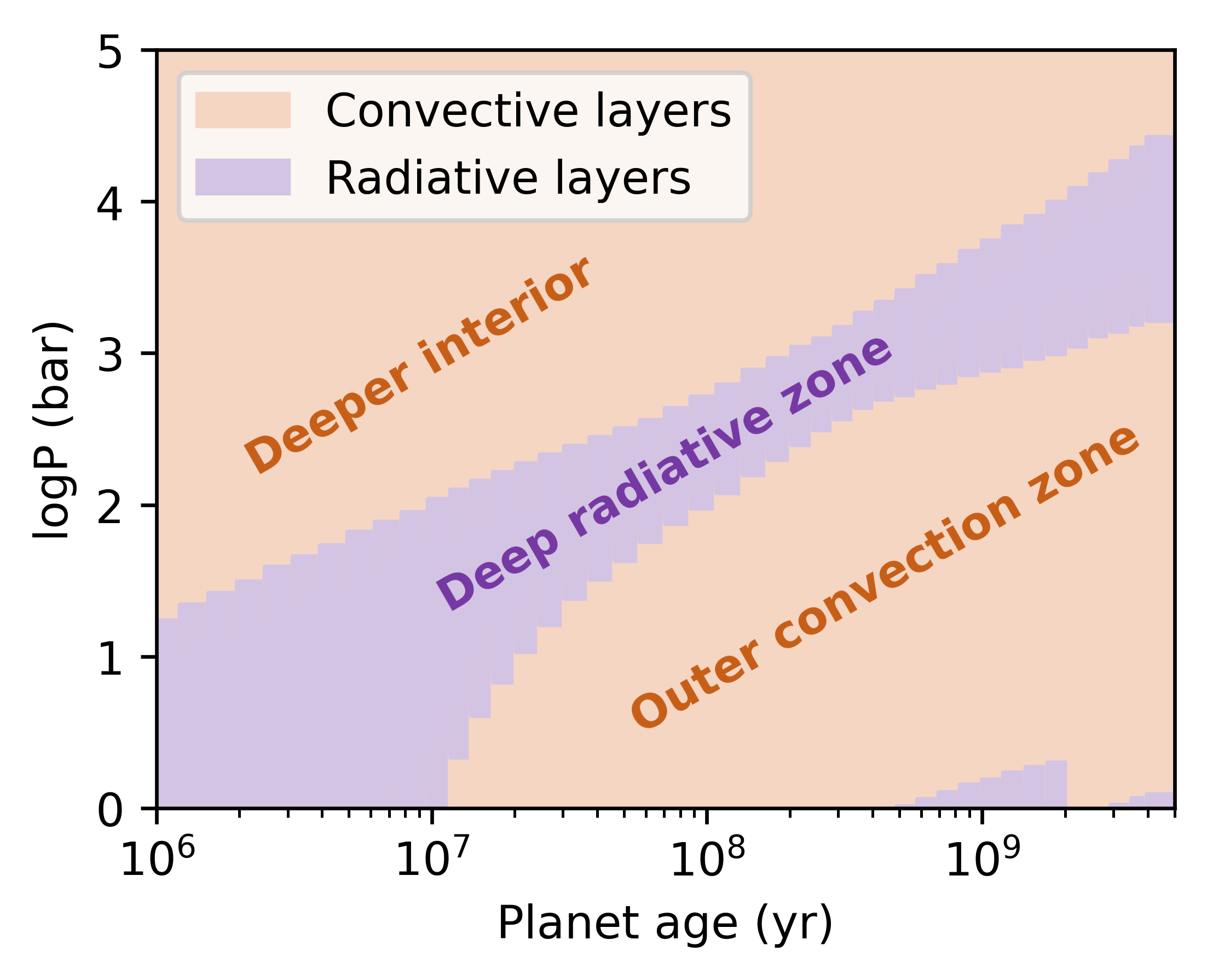}
    \caption{Kippenhahn diagram of the convective (orange) and radiative (purple) layers as Jupiter evolves \addone{for the nominal value of $\alpha = 0.9$ (see \S \ref{sec:methods_opacities})}. \addone{The x-axis is the time after formation}, and the y-axis shows the pressure. For simplicity, we only show the planet above pressures of 1 bar. However, our model includes a radiative photosphere with significantly lower pressures. The lower boundary of the photosphere can be seen transiently around 1 Gyr. The deep radiative zone appears around 10 Myr after formation and is present even today.}
    \label{fig:results_kippenhahn}
\end{figure}

Our results for the \addone{nominal value of $\alpha = 0.9$} are shown as a Kippenhahn diagram in \cref{fig:results_kippenhahn}. The figure starts at a pressure of one bar, similar to the outermost pressure used in interior models. However, our evolution models include a radiative photosphere where the pressure is significantly lower (see \cref{fig:introduction_pie} and the extended Kippenhahn diagrams in Appendix \ref{sec:appendix_kippenhahn_diagrams}). At 1 Myr, there is one large radiative zone in the outer envelope. After around 10 Myr, the radiative zone splits into two: A deep radiative zone caused by the hydrogen-transparency window, and the photosphere at low pressures. As Jupiter evolves, the deep radiative zone moves inward towards higher pressures. Today, the radiative zone is found to be between  \addone{$\sim 1$ and $\sim 50$ kbar}. The location \addone{and appearance} of the radiative region depends on the depth of the hydrogen-transparency window, that is, on the value of $\alpha$. We, therefore, show additional calculations with a range of opacity-scaling factors in Appendix \ref{sec:appendix_kippenhahn_diagrams}. \addone{These additional models suggest that the deep radiative zone exists during the entire evolution when reducing the Freedman opacity by $\sim$ 90\%  at the location of the hydrogen-transparency window. With a reduction of about 80\%, the deep radiative zone is re-established at around 1 Gyr, which would still be early enough to allow for significant enrichment. As discussed previously, similar or stronger opacity reductions can easily be achieved if alkali metals are depleted.}

\addone{Using the nominal value of $\alpha = 0.9$} our results from \cref{fig:results_kippenhahn} suggest that the deep radiative zone appears early during Jupiter's lifetime and persists throughout its evolution. \textbf{We find that the outermost atmosphere stays disconnected from the deeper interior, and therefore they could have different compositions.} However, since this model has a homogeneous envelope, it should be investigated whether the radiative zone is stable against a destabilizing composition gradient, i.e., one where the mean molecular weight decreases inwards. This is investigated in the next section.

\section{Enrichment from above}\label{sec:enrichment_from_above}

In this section, we investigate whether the atmosphere can be enriched from above by accreting high-metallicity objects. We considered two scenarios: Gradual accretion of small heavy-element-rich objects (such as planetesimals, asteroids, or comets) or a single, large impact with a planetary embryo.

\subsection{Accretion of small objects}\label{sec:small_object_accreton}

Here we assume that Jupiter is gradually accreting heavy-element-rich objects during its evolution. \addone{The total accreted heavy-element mass depends on the accretion rate of the objects, their sizes, and densities. The accreted mass is degenerate concerning these quantities: The same outcome can be achieved by either varying the accretion rate or the properties of the properties of the objects. While some estimates of the past and current impact fluxes of small objects for Jupiter exist \citep[see, e.g.,][for recent studies]{2023PSJ.....4..168B,2023PSJ.....4..139N}, the amount of mass accreted by Jupiter after its formation is rather uncertain (see \citet{2018ARA&A..56..137N} for a review). For simplicity, we parametrize the accretion with the total accreted heavy-element mass, ranging from $\sim 10^{-4}$ to $10^{-2} M_\oplus$. This was done by assuming that all objects have the same density and size and using a constant accretion rate of one object every 10 years throughout the evolution.} While the exact composition of the accreted material is important to match the isotopic constraints from observations, here we focus on the total heavy-element enrichment that can be achieved and remain agnostic about the composition. Requirements for the chemical composition of the accreted material are discussed in \S \ref{sec:discussion}.

Another important parameter in these models is the diffusion coefficient $D_{min}$ without large-scale convection or double-diffusive instabilities. To cover a large range of possible vertical transport of chemical species, we allow $D_{min}$ to vary between $10^{-3}$ to $1$ cm$^2$ s$^{-1}$. While it is unclear how deep small objects could penetrate Jupiter's envelope, they should be destroyed long before they reach the deep radiative zone due to the high density of the surrounding gas. \addone{In Appendix \ref{sec:appendix_disruption_location}, we present an order-of-magnitude estimate of the disruption location and show that small objects are disrupted well above the deep radiative zone at any time during the evolution. As a result, the small objects are expected to} deposit all their mass in the outer convection zone (see \cref{fig:introduction_pie,fig:results_kippenhahn}).

\begin{figure}[ht]
    \centering
    \includegraphics[width=\columnwidth]{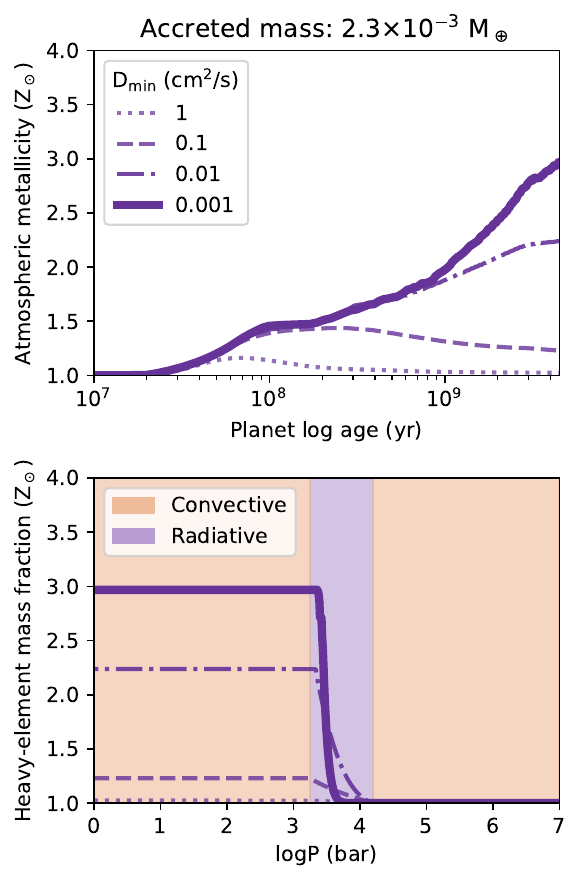}
    \caption{\textbf{Top:} Temporal evolution of the atmospheric metallicity (the heavy-element mass fraction in the outer envelope) in solar units for \addone{the accretion of small objects with} a total accreted mass of $1.6 \times 10^{-3} M_\oplus$, \addone{and using our nominal value of $\alpha = 0.9$}. \textbf{Bottom:} Heavy-element mass fraction in solar units as a function of pressure for Jupiter today. The line styles depend on the assumed $D_{min}$ (see legend). In the bottom panel, the purple and orange shaded regions highlight radiative and convective layers. The radiative zone at a few thousand bar acts as a boundary layer between the outer convective envelope and the deeper interior. Depending on the assumed diffusion through the radiative zone, the atmospheric metallicity can either be very close to or significantly above solar. The thick solid line shows our nominal model with molecular diffusion.}
    \label{fig:results_gradual_stacked}
\end{figure}

First, we present the results for a total accreted mass of $2.3 \times 10^{-3} M_\oplus$. In the top panel of \cref{fig:results_gradual_stacked}, we show how the atmospheric metallicity (the metallicity in the outer convection zone) evolves as a function of time for different $D_{min}$. For $D_{min} = 10^{-3}$ cm$^2$ s$^{-1}$, there is very little material transport through the radiative zone. Almost all the material that is deposited remains in the outer convection zone and does not mix with the deeper interior. This demonstrates that the radiative zone is stable even in the presence of a destabilizing composition gradient. We also find that thermohaline convection does not mix the material through the radiative zone. The nominal model with $D_{min} = 10^{-3}$ cm$^2$ s$^{-1}$, which corresponds to the standard estimated value due to molecular diffusion, leads to an enrichment of about three times solar.

The bottom of \cref{fig:results_gradual_stacked} shows the present-day profiles of the heavy-element mass fractions for the different models. From one to a few thousand bar, the material is homogeneously mixed in the outer convection zone. Then, the radiative zone acts as a boundary layer between the outer and inner convection zones. Depending on $D_{min}$, there is either a steep or gradual decline of the heavy-element mass fraction towards lower pressures.

\begin{figure}[ht]
    \centering
    \includegraphics[width=\columnwidth]{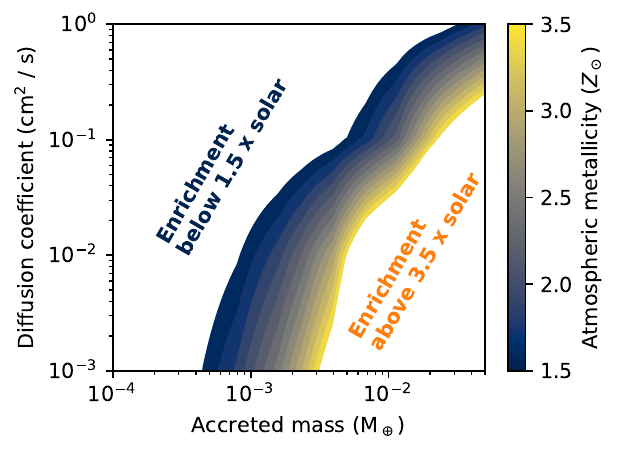}
    \caption{Contours of the atmospheric metallicity in solar units for Jupiter today \addone{for the accretion of small objects and using our nominal value of $\alpha = 0.9$ The x-axis shows the function of the gradually accreted mass, and the y-axis the diffusion coefficient $D_{min}$ in the radiative zone}. If $D_{min}$ is smaller than or equal to the estimated $10^{-2}$ cm$^2$ s$^{-1}$ from molecular diffusion, Jupiter's atmosphere can easily be enriched to two times solar and beyond for a large range of accreted masses. The regions on the left or right of the colored contours either lead to too little or too much enrichment.}
    \label{fig:results_gradual_macc_d_z}
\end{figure}

We next present the results of all the models that we calculated. The details of a few selected final heavy-element profiles are also shown in \cref{fig:appendix_gradual_grid_logP_z} in Appendix \ref{sec:appendix_enrichment_from_above}. \cref{fig:results_gradual_macc_d_z} shows the atmospheric metallicity today as a function of the accreted heavy-element mass and $D_{min}$. A highly enriched outer envelope requires inefficient diffusion with  $D_{min} \leq 10^{-2}$ cm$^2$ s$^{-1}$ or that the accreted mass exceeds a few $10^{-3} M_\oplus$. However, we note that the Galileo measurements and the interior structure models require an enrichment of only about two to three times solar. \textbf{Our results clearly show that this is easily achieved by a large variety of combinations of accreted masses with realistic values for molecular diffusion.}

\subsection{Accretion of a large object}\label{sec:large_object_accretion}

Instead of a gradual accretion of small bodies, as assumed above, an atmospheric enrichment could also be a result of a single collision with a larger object. While no large impacts have been directly observed, theoretical models suggest that they might be common in young planetary systems \citep[e.g.,][]{2015MNRAS.446.1685L}. Here, we investigated whether such an impact can also lead to a stable enrichment of the outer envelope. The simulation setup was similar to what was used in the previous subsection. However, instead of the material being accreted during the entire evolution, we considered an instantaneous impact at a specific time $t_i$. Since the timing of the impact is unknown, we treated $t_i$ as a free parameter ranging from 1 Myr to 1 Gyr. \addone{Note that if $0.7 < \alpha < 0.9$, then the deep radiative zone transiently disappears during the evolution. In those cases, the impact would have to occur after the reappearance of the deep radiative zone, since otherwise the accreted material would be quickly mixed into the interior by large-scale convection.} The enrichment of the outer convection zone (see \cref{fig:introduction_pie}) is determined by how much mass ($m_i$) the impactor loses in these layers. \addone{In Appendix \ref{sec:appendix_disruption_location}, we show that objects with masses less than about $10^{-2} M_\oplus$ should be disrupted before reaching the deep radiative zone about 10 Myr after Jupiter's formation.} We note that $m_i$ does not have to be equal to the mass of the impactor: A bigger object would penetrate much deeper and only lose part of its mass in the outer layers \citep{Liu2019}. Since depositing too many heavy elements at once would be turbulently unstable or lead to an unrealistically high enrichment today, we assumed that the deposited mass $m_i$ ranges from about one to a few times $10^{-3} M_\oplus$. This corresponds to an initial enrichment of about three to nine times solar. The new material was deposited above the deep radiative layer, i.e., at pressures between about 0.1 to 1,000 bar.

\begin{figure}[ht]
    \centering
    \includegraphics[width=\columnwidth]{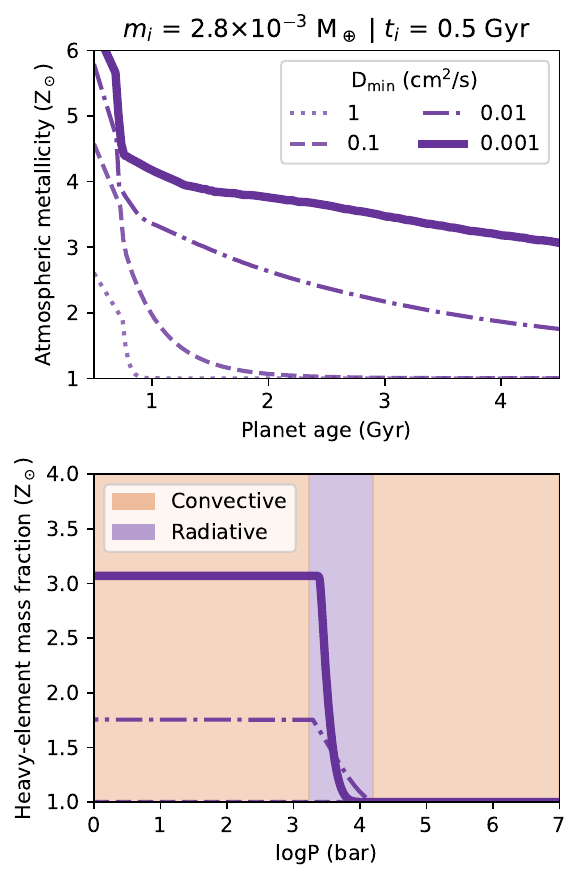}
    \caption{\textbf{Top:} Temporal evolution of the atmospheric metallicity (the heavy-element mass fraction in the outer envelope) in solar units an impact at 500 Myr that deposits of $2.8 \times 10^{-3} M_\oplus$. \addone{These models use our nominal value of $\alpha = 0.9$}.  \textbf{Bottom:} Heavy-element mass fraction in solar units as a function of pressure for Jupiter today. The line styles depend on the assumed $D_{min}$ (see legend). In the bottom panel, the purple and orange shaded regions highlight radiative and convective layers. The radiative zone at a few thousand bar acts as a boundary layer between the outer convective envelope and the deeper interior. Depending on the assumed diffusion through the radiative zone, the atmospheric metallicity can either be very close to or significantly above solar. The thick solid line shows our nominal model with molecular diffusion.}
    \label{fig:results_impact_stacked}
\end{figure}

We first present a case for which (assuming molecular diffusion through the radiative zone), the enrichment at the end of the simulation is about three times solar. This was the case for $m_i = 2.8 \times 10^{-3}$ and $t_i = 0.5$ Gyr. The metallicity in the outer convection zone (atmospheric metallicity) as it evolves with time and the profile today are shown in \cref{fig:results_impact_stacked}. Right after the impact, the outer convection zone is enriched to about six times solar. However, the enrichment drops quickly.  This is not because the material is mixed through the radiative zone by large-scale or thermohaline convection, but because the deep radiative zone moves to higher pressures as time progresses (see \cref{fig:results_kippenhahn}). As a result, the outer convection zone extends into a region with a solar composition, and the heavy-element mass fraction decreases. Afterward, the enrichment slowly decreases due to a combined effect of the diffusion and the expansion of the outer convection zone. Since the enrichment occurs at 0.5 Gyr, there is much time to transport heavy elements via diffusion to the deeper interior, and therefore the results in this case are more sensitive to the assumed material transport compared to the gradual-accretion case (see \cref{fig:results_gradual_stacked} for a comparison). 

\begin{figure}[ht]
    \centering
    \includegraphics[width=\columnwidth]{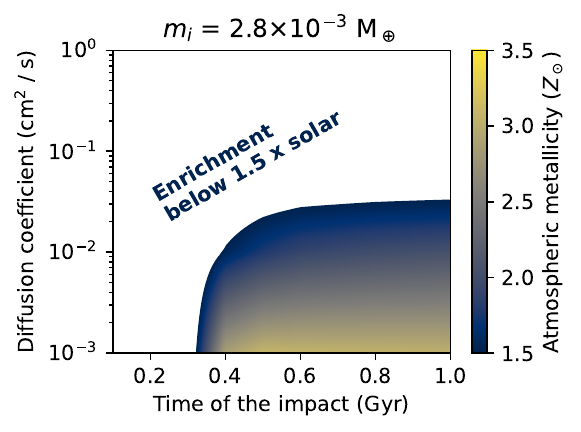}
    \caption{Contours of the atmospheric metallicity in solar units for Jupiter today as a function of the impact time and the diffusion coefficient $D_{min}$ in the radiative zone. In this model, the impactor deposited $m_i = 2.8 \times 10^{-3} M_\oplus$ in the outer convection zone. The regions on top and the left of the colored contours lead to enrichment below 1.5 times solar.}
    \label{fig:results_impact_minD_ti_d}
\end{figure}

\cref{fig:results_impact_minD_ti_d} shows the atmospheric metallicity today for the same deposited mass ($m_i = 2.8 \times 10^{-3} M_\oplus$), but for impact times between 0.1 and 1 Gyr. While larger impacts would be more likely at earlier times, an impact before about 0.4 Gyr does not lead to significant enrichment; at this point the outer convection zone is too shallow. If the impact is later than about 0.5 Gyr, the atmosphere can remain enriched enough to explain the Galileo measurements.

\begin{figure}[ht]
    \centering
    \includegraphics[width=\columnwidth]{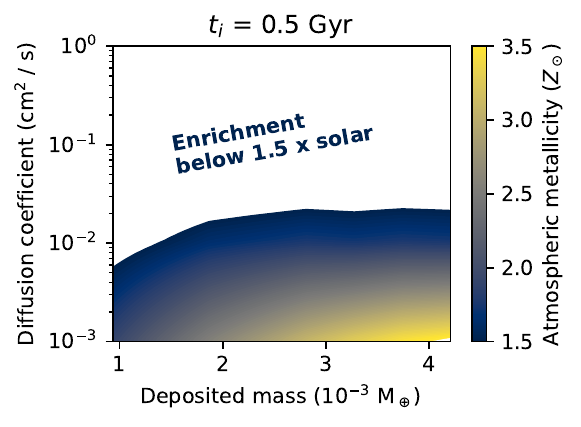}
    \caption{Contours of the atmospheric metallicity in solar units for Jupiter today as a function of the deposited mass by the impactor and the diffusion coefficient $D_{min}$ in the radiative zone. In this model, the impact occurred at 0.5 Gyr. The regions on top of the colored contours lead to enrichment below 1.5 times solar.}
    \label{fig:results_impact_minD_mdep_d}
\end{figure}

\cref{fig:results_impact_minD_mdep_d} shows the atmospheric metallicity today as a function of the deposited mass for $t_i = 0.5$ Gyr. It is clear that unless diffusion is very strong the atmosphere can be significantly enriched if the deposited mass is larger than $\sim 10^{-3} M_\oplus$. For late ($>$ 0.5 Gyr) impacts, the main process that lowers the atmospheric metallicity is diffusion. Therefore, the mass requirement to meet an enrichment of two to three times solar decreases if the impact occurs later. Additional figures for the large-impact scenario are shown in Appendix \ref{sec:appendix_enrichment_from_above}.

Our results demonstrate that the atmosphere can easily be enriched by a large impactor that deposits $\sim 10^{-3} M_\oplus$ heavy elements, as long as the impact does not occur too early. Compared to the gradual-accretion scenarios, it is more difficult to have very high atmospheric metallicities ($>$ 3.5 times solar) because diffusion has more time to transport the heavy elements through the deep radiative zone into the deeper interior.

\section{Could Jupiter's atmospheric enrichment come from below?}\label{sec:enrichment_from_below}

Giant planet formation models have shown that Jupiter likely had composition gradients in the interior as a result of planetesimal-envelope interactions \citep[e.g.,][]{Lozovsky2017,Helled2017,Valletta2019,Stevenson2022}. Heavy elements could diffuse out of a compact or dilute core \citep{Guillot2004,Moll2017}, or convective mixing \citep{Vazan2018,Mueller2020} could erode the core. Large-scale convection then carries the heavy elements into the upper envelope. 

\addone{This scenario cannot resolve the mismatch between the interior models and the atmospheric measurements: Enriching the entire envelope disagrees with the gravitational moments and structure models, because they require an under-density in the envelope compared to a two-times solar composition \citep[e.g.,][]{Debras2019,2022A&A...662A..18M,2023arXiv230209082H}. Also, in this case, a deep radiative zone can't exist, since it would prevent the core material from reaching the atmosphere. Instead, the core material would be stuck beneath the radiative zone. 
Transport of heavy elements from the core to the atmosphere would only be possible if the radiative zone disappears and reappears during Jupiter's evolution (see Appendix \ref{sec:appendix_kippenhahn_diagrams}).}

\addone{There are also other potential issues: Compared to enriching the atmosphere from above, doing it from below requires a lot of heavy elements since the metallicity of the entire envelope has to be raised. The primordial dilute core must therefore be significantly eroded. This is inconsistent with current interior structure models of Jupiter, which require a large dilute core \citep{Wahl2017,2021PSJ.....2..241N,2022PSJ.....3...89I,2022A&A...662A..18M,2022PSJ.....3..185M,2023arXiv230209082H}. Nevertheless, it has been suggested that the observed elemental abundances in Jupiter's atmosphere are partly the result of eroding this primordial dilute core \citep[e.g.,][]{2019AJ....158..194O}. Therefore, it is valuable to explore whether the atmosphere can be enriched from below.}

We investigate the enrichment-from-below hypothesis by evolving post-formation Jupiter models that include a dilute core. The thermal state of proto-Jupiter and the shape of the composition gradient largely determines the amount of mixing that would occur during the evolution \citep[e.g.,][]{Vazan2015}. Therefore, we considered a large range of post-formation models of Jupiter. The models were constructed by combining different initial thermal states (parameterized by the initial central temperature) and three different composition gradients (see \cref{fig:appendix_dilute_core_initial_conditions}). The initial atmospheric metallicity was always set to solar. \addone{In these models (for the reason discussed earlier) we did not impose an opacity window at a few 1000 K and used the Freedman radiative opacity at low temperatures.}

\begin{figure}[ht]
    \centering
    \includegraphics[width=\columnwidth]{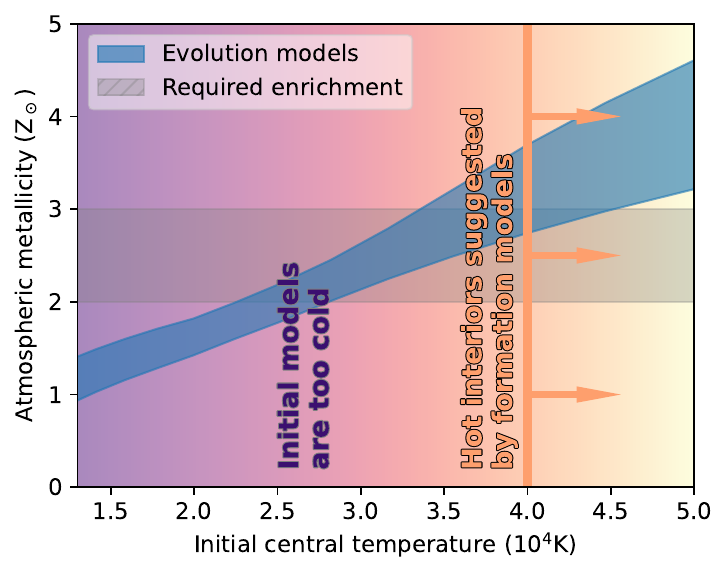}
    \caption{Atmospheric metallicity in solar units vs. initial central temperature. The shaded blue region shows the evolution models. To match the required enrichment (shaded gray region) from measurements, the initial models must be unrealistically cold. Formation models suggest that Jupiter forms much hotter: This is indicated by the orange line and the arrows pointing towards higher initial central temperatures. These models lead to a too-high atmospheric metallicity. See text for details.}
    \label{fig:results_dilute_core_atmospheric_metallicity}
\end{figure}

For all combinations of the initial conditions, we simulated Jupiter's evolution (including mixing by large-scale- or semi-convection) until today. Depending on the combination of the initial thermal state and composition gradients, the outcomes ranged from mostly intact dilute cores to fully homogeneous envelopes. \cref{fig:results_dilute_core_atmospheric_metallicity} shows the atmospheric metallicity as a function of the initial central temperature. All the results were combined to create the shaded blue region, which shows the range of possible outcomes. We find roughly a linear relationship between Jupiter's atmospheric metallicity today and the initial temperature.

It is clear that for the evolution models to match Jupiter's atmospheric metallicity of about two or three times solar, Jupiter could not have formed with central temperatures below about 40,000 K. This currently appears unlikely, because formation models suggest that Jupiter formed hotter than that \citep{Cumming2018,Valletta2020,2022PSJ.....3...74S}. Alternatively, the erosion of the dilute core by large-scale convection must have been otherwise inhibited. If that were true, however, then the envelope could not have been significantly enriched with core material. Consequently, from a formation and evolution point of view, it is difficult to obtain the required atmospheric enrichment.

Overall, we conclude that the enrichment-from-below hypothesis is improbable because it contradicts formation-evolution models, Jupiter's atmospheric enrichment, and its gravitational moments.

\section{Discussion}\label{sec:discussion}
The models presented in this work use a deep radiative zone to keep the atmosphere enriched down to pressures of about 1,000 bar. Such a radiative region was first presented in \citep{1994Icar..112..354G}, where a hydrogen-transparency window was suggested as the cause. This idea recently received significant support \citep{2023NatAs...7..678C}. While our nominal models use the location and depth of the opacity window from \citep{1994Icar..112..354G}, in Appendix \ref{sec:appendix_kippenhahn_diagrams}, we present additional models that show how the depth of the opacity window affects the appearance of the deep radiative zone. 

\addone{The deep radiative zone persists throughout the entire evolution only when a significant reduction of the opacity (by 90\%) is assumed. Such an opacity reduction is expected if alkali metals are depleted \citep{Freedman2008}, as seems to be the case for Jupiter \citep{2023ApJ...952L..27B}. A smaller reduction, of less than 80\% can cause the radiative zone to disappear. This would (transiently) make the entire envelope convective, and early accretion would lead to the enrichment of the deeper interior. In this case, any present-day enrichment would have to be accreted once the radiative zone reappears. This requires higher accretion rates at later times (beyond $\sim$ 1 Gyr) to achieve an enrichment of a few times solar. However, 1 Gyr is early enough for a significant enrichment.}

\addone{The exact location of the radiative zone could vary depending on the details. If the radiative zone occurs higher up in the envelope less mass would be needed to enrich the atmosphere. Therefore, while our results were calculated assuming a specific mechanism to create the deep radiative zone, another mechanism would work as well. While different parameters would be required to enrich the atmosphere, the results would be similar.}

We also note that the deeper interior in our models was cooling adiabatically. If the deeper interior were (partially) non-adiabatic, for example, because of a large dilute core or helium rain, the planet's cooling would be affected \citep{Leconte2012,Mankovich2016,2020ApJ...889...51M}. This could, depending on the opacity, change the timing of when the deep radiative zone appears or disappears. \addone{An extended dilute core in Jupiter inhibits large-scale convection in the deep interior and therefore slows down the heat transport. However, the location of the deep radiative zone is unlikely to be affected by it since it is mostly affected by the opacity in the outer region. Similarly, the process of helium rain is also unlikely to change our conclusions, since evolution models that include hydrogen-helium demixing show that it would only occur after a few billion years \citep[e.g.,][]{Mankovich2016,2020ApJ...889...51M}. Compared to the large uncertainties on the opacity and the mixing, these are  minor effects. Nevertheless, it would be desirable for future studies to include as many processes as possible and investigate their interplay and their effect on Jupiter's long-term evolution.} 

The second important mechanism in our models is the vertical transport of chemical species through the deep radiative zone. Estimates for the diffusion in the Earth's stratosphere \citep{1981JGR....86.9859M,1998JFM...375..113D} and Jupiter's upper atmosphere \citep{2016Icar..276...21W} exist. However, the diffusion at a depth of $10^3$ to $10^4$ bar where the deep radiative zone lies is highly uncertain. While for stellar interiors a full treatment of temperature, pressure, and chemical diffusion exists \citep[e.g.,][]{1969fecg.book.....B,1986ApJS...61..177P,1994ApJ...421..828T}, this is currently unavailable for the conditions relevant to the atmospheric enrichment of Jupiter.  Additional measurements and simulations would be required to fully determine the transport properties under these conditions. Given these uncertainties, we intentionally kept our treatment of diffusion straightforward. The transport of elements was modeled as a simple vertical transport through the deep radiative zone parameterized by a minimal diffusion coefficient \addone{spanning several orders of magnitude}. A reasonable assumption is that this coefficient would be similar to the molecular diffusion coefficient $D \sim 10^{-3}$ cm$^2$/s, which we have used in our nominal models. However, we also showed that even if the vertical transport is about an order of magnitude more efficient ($D \sim 10^{-2}$ cm$^2$/s,) the gradual-accretion scenario can still enrich the atmosphere to two to three times solar for reasonable accretion rates.

\addone{Other fluid dynamic instabilities could also be present, leading to an increased eddy diffusivity compared to molecular diffusion. Our results suggest that, for example, a diffusion coefficient of $D = 1$ cm$^2$/s would require a significant accretion of heavy elements which is probably unrealistic (see \cref{fig:results_gradual_macc_d_z,fig:results_impact_minD_mdep_d}. Therefore, to assess whether the atmosphere can stay enriched over evolutionary timescales, constraining the vertical transport of species through the radiative zone is crucial. We encourage future investigations of this topic.}

\citet{2023A&A...680L...2H} recently investigated the scenario of an enriched atmosphere in Jupiter with an inverted heavy-element gradient. Two scenarios in which there is a dynamically stable layer at $\sim 10^3$ or $\sim 10^6$ bar were considered. The former location corresponds to a radiative layer due to the locally decreased opacity from \citet{1994Icar..112..354G} and the latter to the helium-rain region in Jupiter. The interior and evolution of Jupiter were simulated to assess whether these scenarios would allow for an inverted heavy-element gradient. Because the stable layer due to helium rain is at high pressures, \citet{2023A&A...680L...2H} suggest that the mass required for the enrichment is too large. It was found that the enrichment at lower pressures would be stable and satisfy the current gravity field constraints. However, it was argued that this scenario is unlikely due to the diffusion through the radiative zone, the mass budget required, and the isotopic constraints. 

Our results are compatible with those from \citet{2023A&A...680L...2H} in the sense that the deep radiative zone is stable against an inverted heavy-element gradient. However, our conclusion is different regarding the feasibility of this scenario: While \citet{2023A&A...680L...2H} performed evolution calculations to investigate the appearance of the deep radiative zone, these evolution models, unlike ours, did not include the accretion or mixing of heavy elements. \addone{Using results from a recent dynamical study of the primordial Kuiper belt \citep{2023PSJ.....4..168B}, they also compared the available mass through collisions with that required for an atmospheric enrichment. \citet{2023A&A...680L...2H} suggest that in the first 500 Myr about $2 \times 10^{-3} M_\oplus$ could be accreted. Since the collision rate is decreasing with time, this is about an order of magnitude larger than what could be accreted in the last Gyr. To satisfy the elemental and isotopic abundances of Jupiter's atmosphere \citep[e.g.,][]{2022arXiv220504100G}, they argue that the material should be accreted early.}

\addone{In this study, we investigated the accretion scenario in much more detail. We considered different accretion scenarios as well as mixing processes using an advanced planetary evolution model, and assuming a large range of parameters. Our results clearly show that reasonable values for the accreted mass and the vertical transport through a deep radiative zone in Jupiter can easily lead to an enrichment of two to three times solar in Jupiter today. This provides a new view on the connection between the atmospheric composition and bulk composition of Jupiter, and giant planets in general.}

\addone{We note that Jupiter has protosolar $^{15}$N/$^{14}$N and D/H isotopic ratios. In some of the scenarios presented in this work, the material accreted after the first $\sim$ 10 Myr of Jupiter's evolution could introduce a deviation from protosolar ratios. Such elemental and isotopic constraints could be used to further constrain formation and evolution models and we hope to address it in future research. However,  it should be kept in mind that for giant planets there are large uncertainties in the composition of the accreted planetesimals \citep{2021ApJ...909...40T,2022MNRAS.509..894H}, pebbles \citep{2017MNRAS.469.4102M,2017MNRAS.469.3994B,2019MNRAS.487.3998B,2021A&A...654A..71S,2021A&A...654A..72S} and gas \citep{2022ApJ...937...36P}, and therefore linking the current composition of a planet with its formation path is extremely challenging. For Jupiter, various studies have shown that the expected atmospheric composition also depends on the details of the planet's formation pathway \citep{2021A&A...651L...2O, 2021ApJ...918L..23M,2022PSJ.....3..141A}. In any case, a better understanding of the chemical makeup of Jupiter's atmosphere and a determination of the composition of a broad range of small solar-system objects is crucial in constraining the potential enrichment pathways \citep{Lunine2023late}.}

\addone{As we show in \S \ref{sec:large_object_accretion}, if the deep radiative zone is established early} it is also plausible that the enrichment occurred quite early in Jupiter's history, and the composition of the impactor could be different from what we observe in the solar system today. \addone{Furthermore, chemical processes occurring in Jupiter's atmosphere and upper envelope could affect the observed composition \citep{2004jpsm.book...59T}; for example, the altering of post-impact molecular abundances due to the recycling of chemical species should be considered \citep{2010Icar..209..602V}.}

Finally, while this study focused on Jupiter, a similar mechanism could operate in the other solar-system giants (Saturn, Uranus, and Neptune) as well as giant exoplanets. In particular, this would have crucial implications for the interpretation of the measured atmospheric compositions of exoplanets by, e.g., the James Webb Space Telescope \citep{Gardner2006,2015MNRAS.448.2546B} or the Ariel mission \citep{Tinetti2018}. It is clear that understanding both the atmospheric and bulk compositions of giant planets is essential to constrain planetary origins \citep[e.g.,][]{2014PNAS..11112601B,Teske2019,2021ApJ...909...40T,2022arXiv221100649E}. However, our results suggest that the measured atmospheric metallicity of giant planets does not have to represent their bulk composition. Therefore, to be able to interpret current and future measurements, it is essential to better understand the connection between the atmospheres and the deep interiors of giant planets\addone{, starting with Jupiter}

\addone{There is a clear need for more data and improved simulations to better constrain whether there is a deep radiative zone on Jupiter, and if so, determine its origin and location. This will require a better understanding of opacities in giant planets, including the potential contributions of grains or clouds. To further constrain the pathways of atmospheric enrichment, a better understanding of the vertical mixing through the radiative-convective boundary is needed. Additionally, improved hydrodynamic simulations to investigate mixing processes including double-diffusive instabilities would help to improve evolution models.}

\addone{Finally, a Saturn probe would allow us to do a comparative analysis of Jupiter's and Saturn's atmospheres and identify whether also for Saturn there is tension between the atmospheric composition and interior models. Similarly, accurate measurements of the atmospheric composition of a large number of giant exoplanets, together with estimates of the planetary bulk composition would provide a more global view of the relation between the atmospheric and bulk compositions of giant planets \citep{Helled2022,Muller2023}.}


\section{Conclusions}\label{sec:conclusions}

In this paper, we modeled Jupiter's evolution to investigate possible pathways to enrich the planet's outer envelope. It has been suggested previously that Jupiter has a deep radiative zone that disconnects the outer envelope from the deeper interior. Here, we investigated whether such a radiative zone could be present during Jupiter's evolution and whether it would be stable enough to prevent the downward mixing of heavy elements. We considered two main scenarios: Enrichment from above by accretion of small or large objects, or from below by core erosion. For both cases, we have performed extensive simulations to cover a large range of parameters.

Our most important results are:

\begin{itemize}
    \item \addtwo{A local reduction of the opacity (an opacity window) at $\sim$ 2000 K by $\sim$ 90\%} \addone{due to hydrogen transparency or depletion of alkali metals creates a deep radiative zone that appears shortly after Jupiter's formation and can persist until today.}
    \item The deep radiative zone in Jupiter is stable even if there is a destabilizing composition gradient.
    \item Gradual accretion of small heavy-element-rich objects or a single collision with a large object can easily enrich Jupiter's atmosphere to two or three times solar metallicity and above. \addone{If the mixing through the radiative zone is governed by molecular diffusion (with $D \lesssim 10^{-2}$ cm$^2$/s), the enrichment persists.}
    \item The erosion of a primordial dilute core is unlikely to explain Jupiter's atmospheric enrichment.
    \item Our results strongly imply that atmospheric composition measurements of exoplanets do not have to be representative of the bulk envelope.
\end{itemize}

Our results suggest that it is possible that Jupiter's atmospheric enrichment does not represent the metallicity of its deeper interior. This possibility, if correct, would resolve the mismatch and long-standing problem of the tension between the measured atmospheric enrichment and the results from interior models of Jupiter. 

In addition, these results imply that measurements of the atmospheric composition of exoplanets should be interpreted with great caution. The atmospheric metallicity of giant exoplanets can not only significantly differ from the deep interior, but it could also be higher than the bulk metallicity.

\begin{acknowledgments}
    We thank D. Stevenson, J. Lunine, S. Howard and T. Guillot for fruitful discussions. We also thank S. Markham for deriving the estimate of the molecular diffusion coefficient, \addone{and the anonymous reviewer for helpful comments.} We also thank the Juno  Science team, in particular Juno's Interior and Origins Working Groups.  Finally, we acknowledge support from SNSF grant \texttt{\detokenize{200020_188460}} and the National Centre for Competence in Research ‘PlanetS’ supported by SNSF.
\end{acknowledgments}

\software{
MESA \citep{Paxton2011,Paxton2013,Paxton2015,Paxton2018,Paxton2019,2022arXiv220803651J},
NumPy \citep{harris2020array},
SciPy \citep{2020SciPy-NMeth},
Matplotlib \citep{Hunter2007},
Jupyter \citep{jupyter}
}

\clearpage
\appendix

\section{The influence of the opacity scaling factor}\label{sec:appendix_kippenhahn_diagrams}

\begin{figure}[ht]
    \centering
    \includegraphics[width=0.9\columnwidth]{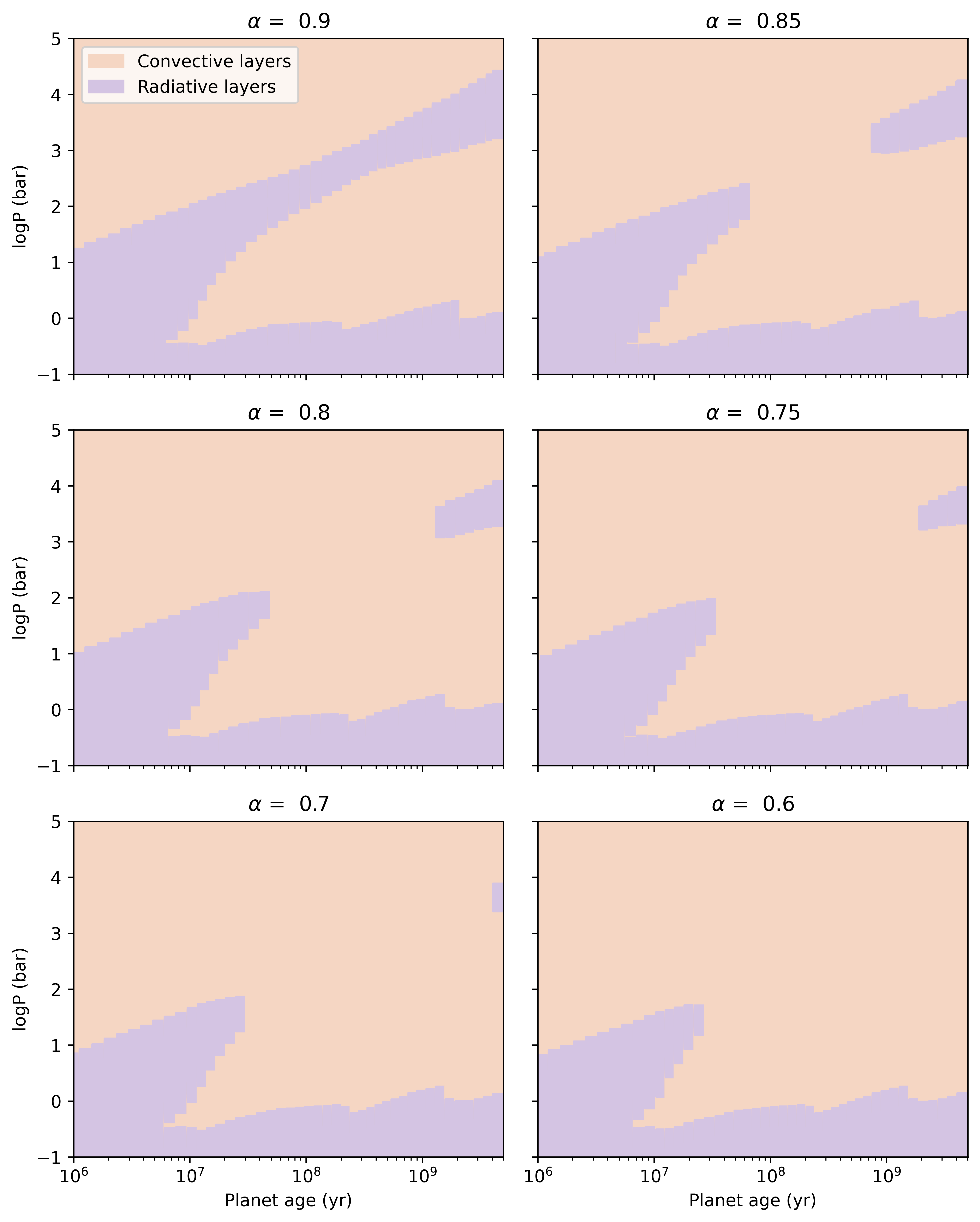}
    \caption{Kippenhahn diagrams of the convective (orange) and radiative (purple) layers as Jupiter evolves. The x-axis is the time after formation, and the y-axis shows the pressure. From the bottom up, the photosphere, outer convection zone, deep radiative zone, and deeper interior can be seen. Each panel uses a different opacity scaling factor $\alpha$ (see \cref{eq:opacity_scaling}). For every model, the deep radiative zone appears around 10 Myr. Depending on $\alpha$, the deep radiative zone can disappear after a few 10 Myr, but then also reappear later. For $\alpha \leq 0.6$ there is no deep radiative zone today.}
    \label{fig:appendix_kippenhahn_logP_grid}
\end{figure}

\begin{figure}[ht]
    \centering
    \includegraphics[width=0.9\columnwidth]{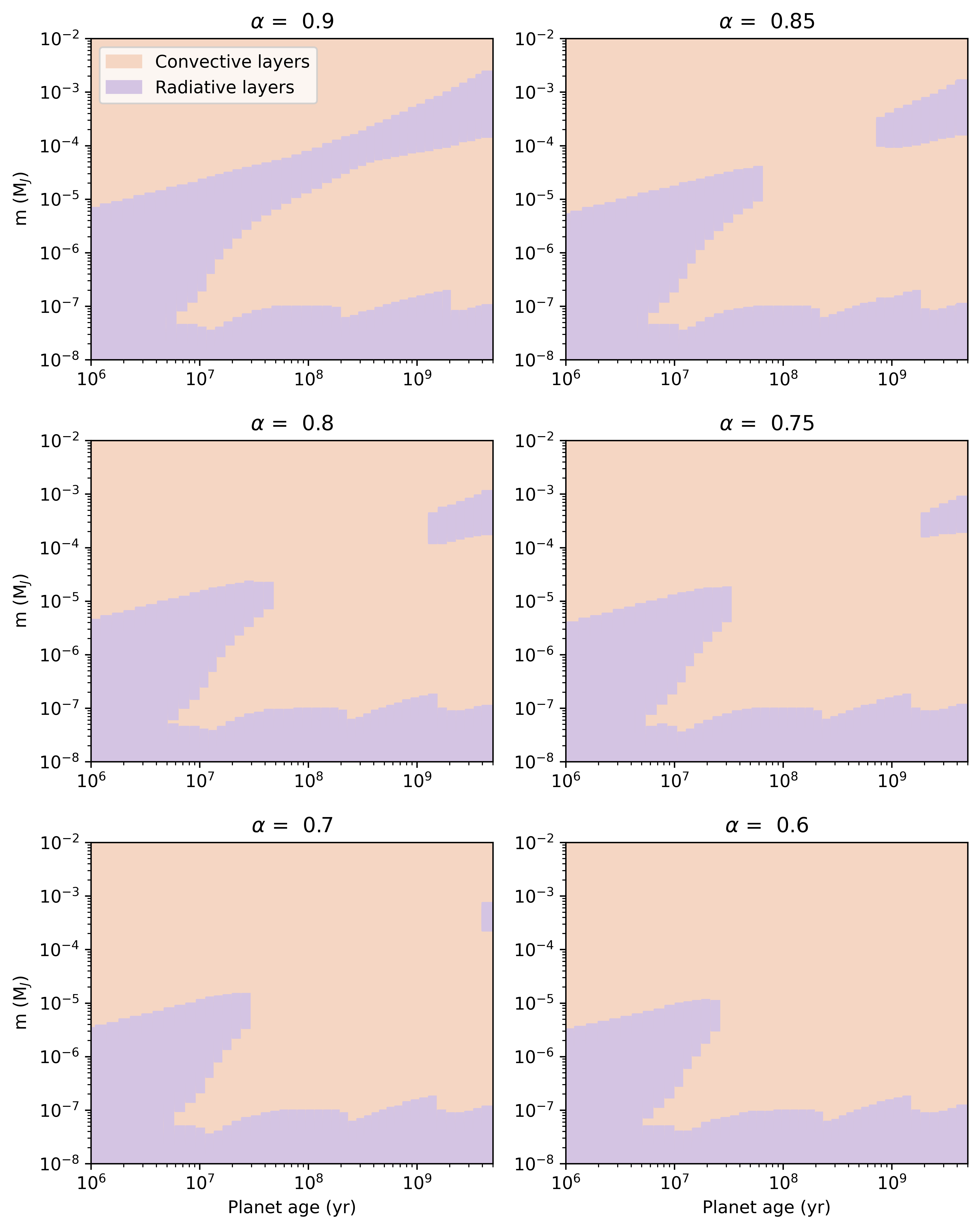}
    \caption{Similar to \cref{fig:appendix_kippenhahn_logP_grid}, but the y-axis shows the mass coordinate instead. The mass coordinate is defined from the outside-in, with $m = 0$ at the surface and 1 M$_J$ at the center.}
    \label{fig:appendix_kippenhahn_q_grid}
\end{figure}

In \S \ref{sec:radiative_zone} we showed that our nominal models lead to a deep radiative zone that appears early and stays throughout Jupiter's evolution. To better understand how this behavior depends on the depth of the opacity window, we performed additional simulations using different values for the opacity scaling factor $\alpha$ (see \S \ref{sec:methods_opacities}). The Kippenhahn diagrams of these simulations are shown in  \cref{fig:appendix_kippenhahn_logP_grid,fig:appendix_kippenhahn_q_grid}, where the y-axis uses the pressure or normalized mass coordinate. If $\alpha < 0.9$, the deep radiative zone disappears during the evolution, usually after a few 10 Myr. Then, depending on the value of $\alpha$, it reappears after about one to a few Gyr. This is not the case for $\alpha \leq 0.6$: The deep radiative zone is only present during the first $\sim$ 10 Myr and then it vanishes for good. For the deep radiative zone to be present for most of Jupiter's lifetime, there is a minimum depth of the opacity window. These models suggest that as long as $\alpha \gtrsim 0.8$, a gradual enrichment of the atmosphere is possible. For the scenario of accreting a larger impact, the impact needs to occur after the re-appearance of the deep radiative zones.

\section{Estimating the disruption location}\label{sec:appendix_disruption_location}

\addone{Here, we do an order-of-magnitude estimate of whether impactors in the mass range that we consider are disrupted above the deep radiative zone. We follow \citet{2018ApJ...864..169J} to estimate the disruption radius of an object entering the atmosphere of Jupiter. \addtwo{We define disruption as the first breakup stage. For simplicity, we do not consider the further breakup into increasingly smaller (fragmentation).} An impactor is disrupted when the integrated ram pressure over the cross-section of the impacting object approaches its binding energy.  This assumption leads to a simple criterion for disruption:}

\begin{equation}
    f \equiv \frac{\rho_{r} v^2}{\bar{\rho_i} v_{esc, i}^2} \approx 1
    \label{eq:disruption}
\end{equation}

\addone{where $\rho(r)$ is the density in the atmosphere at the radial location $r$, $v_i$ is the velocity and $\bar{\rho_i}$ the mean density of the impactor. The escape velocity at the surface of the impactor is $v_{esc, i}^2 = G M_i / R_i$, with $M_i$ and $R_i$ the mass and radius of the object. The velocity $v_i$ at the time of disruption should be approximately the orbital velocity at the location of Jupiter $(G M_* / a_{orb})$.}

\begin{figure}
    \centering
    \includegraphics[width=0.6\columnwidth]{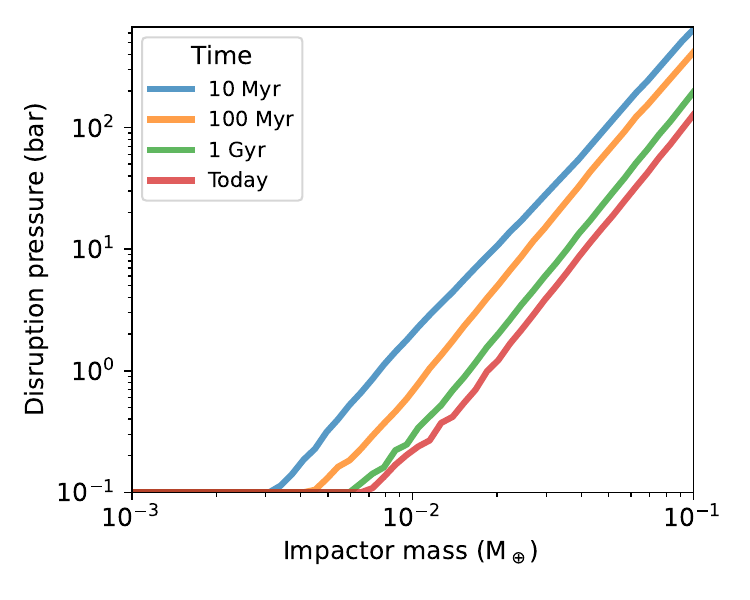}
    \caption{Disruption pressure as a function of the mass of the impactor for four different times during Jupiter's evolution. \addtwo{Disruption is defined as the first breakup stage, which does not necessarily correspond to the airburst height of the larger fragments.} Since the outer boundary in the evolution model is at 0.1 bar, this is the lowest possible disruption pressure. Therefore, the disruption pressure is constant for low-mass objects. Without this limitation, they would be disrupted at lower pressures.}
    \label{fig:appendix_disruption}
\end{figure}

\addone{Since Jupiter is cooling and getting denser with time, the disruption location is time-dependent. To estimate where the impactor is disrupted, we solve \cref{eq:disruption} for $\rho(r)$ from our models at different times during the evolution, and get the disruption radius and pressure. We calculate these quantities for masses between $M_i = 10^{-3}$ to $10^{-1} M_\oplus$, and use a density of $\bar{\rho_i} = 1$ g/cm$^3$ to estimate $R_i$. The disruption pressures as a function of impactor mass are shown for Jupiter at 10, 100, 1000 Myr, and today in \cref{fig:appendix_disruption}. At low masses, the disruption pressure is constant at 0.1 bar. This is an artifact of our models since that is the pressure at the outer boundary of the model. Without this limitation, the low-mass impactors should be disrupted at similar or lower pressures. This is also true for the low-mass objects that we consider in this work (see \S \ref{sec:small_object_accreton}). \addtwo{We note that after disruption the larger fragments may still penetrate deeper. Therefore, depending on the size of the fragment (and its material strength), the disruption location may be different from the airburst height at which the object is destroyed.} Impact studies have shown that the largest fragment of Shoemaker-Levy 9 \addtwo{exploded} at a pressure of about 2 bar \citep{1996EM&P...73..147H}. \addtwo{This suggests that low-mass objects can be destroyed at much shallower pressures than the location of the deep radiative zone.}}

\section{Additional enrichment-from-above models}\label{sec:appendix_enrichment_from_above}

\begin{figure}[ht]
    \centering
    \includegraphics[width=\columnwidth]{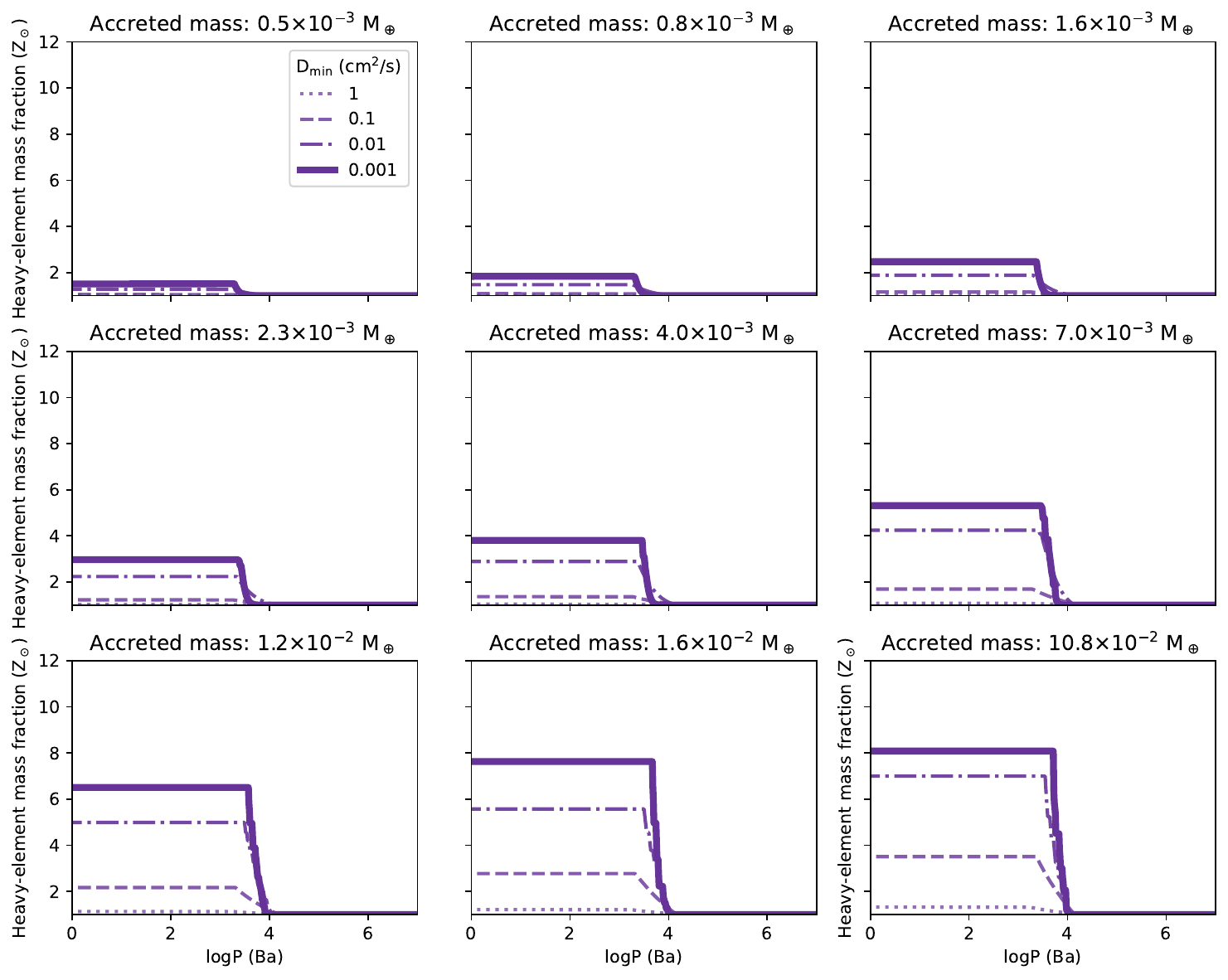}
    \caption{Heavy-element profile as a function of pressure for Jupiter today for the gradual-accretion models. Each panel shows a model with a different total accreted mass. The color shades correspond to the diffusion coefficient in the radiative zone (see legend).}
    \label{fig:appendix_gradual_grid_logP_z}
\end{figure}

\begin{figure}[ht]
    \centering
    \includegraphics[width=\columnwidth]{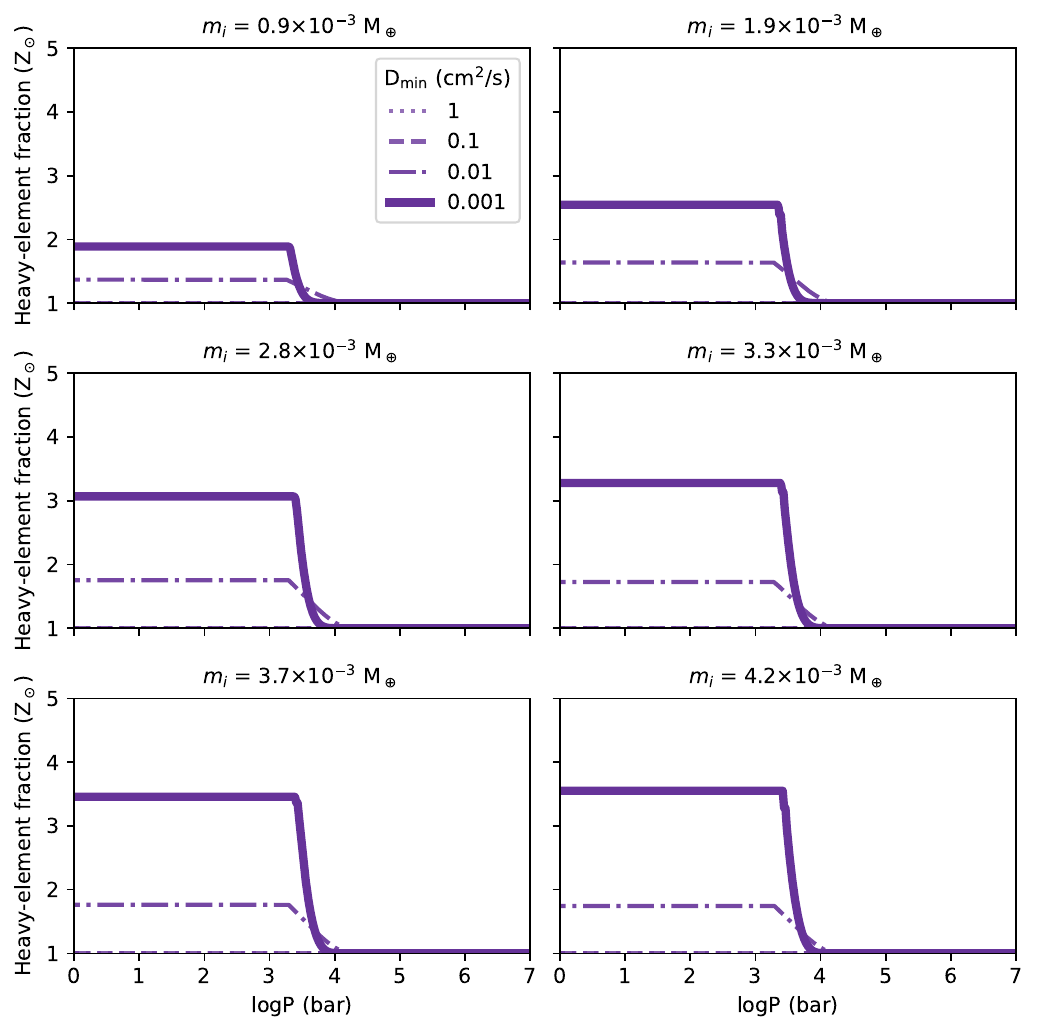}
    \caption{Heavy-element profile as a function of pressure for Jupiter today for the gradual-accretion models. Each panel shows a model with a different total accreted mass. The color shades correspond to the diffusion coefficient in the radiative zone (see legend).}
    \label{fig:appendix_impact_grid_logP_z}
\end{figure}

\begin{figure}[ht]
    \centering
    \includegraphics[width=\columnwidth]{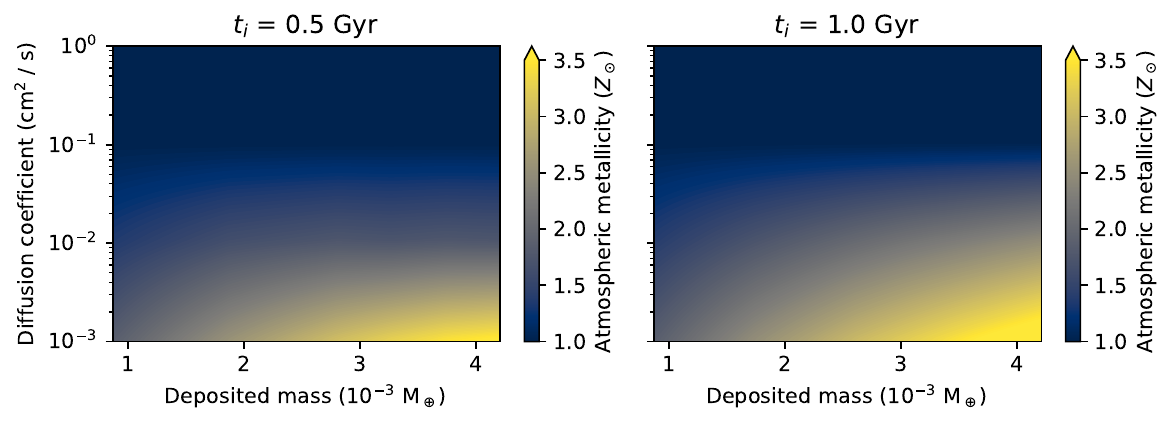}
    \caption{Contours of the atmospheric metallicity in solar units for Jupiter today as a function of the deposited mass by the impactor and the diffusion coefficient $D_{min}$ in the radiative zone. The two panels show different impact times.}
    \label{fig:appendix_impact_contours_m_grid}
\end{figure}

\cref{fig:appendix_gradual_grid_logP_z} and \cref{fig:appendix_impact_grid_logP_z} show the heavy-element profiles today as a function of pressure for a few selected gradual-accretion and large-impact models. In \cref{fig:appendix_impact_contours_m_grid}, we show a more detailed version of \cref{fig:results_impact_minD_mdep_d} from the main text: Here, we also show the case of an impact at 1 Gyr. This comparison shows that the timing of the impact becomes less important once the deep radiative zone arrives at higher pressures. It still requires less mass to enrich the outer convection zone if the impact is late because diffusion has less time to mix the material with the deep interior. However, the mass-requirement difference between an impact at 0.5 or 1 Gyr is marginal.

\section{Initial conditions for the enrichment-from-below scenario}\label{sec:appendix_enrichment_from_below}

\begin{figure}[ht]
    \centering
    \includegraphics[width=1\columnwidth]{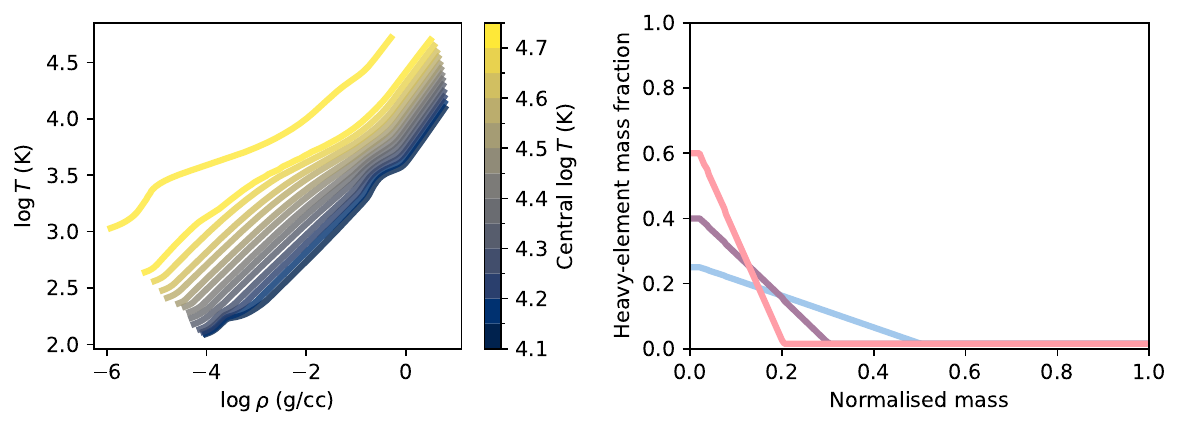}
    \caption{Density-temperature (left) and composition (right) profiles used as initial conditions for the enrichment-from-below scenario in \S \ref{sec:enrichment_from_below}.}
    \label{fig:appendix_dilute_core_initial_conditions}
\end{figure}

\cref{fig:appendix_dilute_core_initial_conditions} shows the density-temperature and the composition profiles that were used as initial conditions for the models presented in \S \ref{sec:enrichment_from_below}. The initial density-temperature profiles were calculated using an adiabatic hot-start model that was allowed to cool for a certain amount of time. The composition gradients, inspired by formation and interior models of Jupiter \citep[see, e.g.,][for a review]{2022Icar..37814937H}, were then imposed using the \textit{relax\_initial\_composition} method available in MESA. We note that current formation models generally suggest that Jupiter should form very hot \citep{Cumming2018,Valletta2020,2022PSJ.....3...74S}. Here, to explore a range of possible pathways, we consider a much broader range of initial thermal states.


\clearpage
\bibliography{library}
\bibliographystyle{aasjournal}

\end{document}